\documentclass{PHYEAUTH}
\usepackage{graphicx}
\begin{document}
\begin{frontmatter}
\title{Novel Electronic States in Graphene Ribbons\\
-Competing Spin and Charge Orders-}
\author[address1]{Atsushi~Yamashiro},
\author[address1,address2]{Yukihiro~Shimoi},
\author[address1,address2]{Kikuo~Harigaya\thanksref{thank1}},\\
and
\author[address3]{Katsunori~Wakabayashi} 

\address[address1]{Nanotechnology Research Institute, National Institute of Advanced
Industrial Science and Technology (AIST),\\ 1-1-1 Umezono, Tsukuba 305-8568, Japan}
\address[address2]{Research Consortium for Synthetic Nano-function Materials Project (SYNAF), National Institute of Advanced Industrial Science and Technology (AIST), 1-1-1 Umezono, Tsukuba 305-8568, Japan}
\address[address3]{Department of Quantum Matter Science, Graduate School of Advanced Sciences of Matter (ADSM),\\ Hiroshima University,  Higashi-Hiroshima 739-8530, Japan}
\thanks[thank1]{
Corresponding author. 
E-mail: k.harigaya@aist.go.jp}

\begin{abstract}
\  In a nanographene ring with zigzag edges, the spin-polarized  state and the charge-polarized state are stabilized by the on-site and the nearest neighbor Coulomb repulsions, $U$ and $V$, respectively, within the extended Hubbard model under the mean field approximation.
In a M\"{o}bius strip of the nanographene with a zigzag edge, $U$ stabilizes two magnetic states, the domain wall state and  the helical state. Both states have ferrimagnetic spins localized along the zigzag edge while the  former connects the opposite ferrimagnetic orders resulting in a magnetic frustration forced by the topology  and the latter rotates the ferrimagnetic spins uniformly to circumvent the frustration. The helical state is lower in energy than the domain wall state. On the other hand, $V$ stabilizes another domain wall state connecting the opposite charge orders. 
\end{abstract}

\begin{keyword}
nanographite \sep edge state \sep M\"{o}bius strip\sep helical magnetism\sep domain wall \sep extended Hubbard model
\PACS 71.10.Hf \sep 73.22.-f  \sep  73.20.At \sep 75.75.+a
\end{keyword}
\end{frontmatter}

%[main text]
\section{Introduction}
Nanoscale materials are strongly affected by their geometries.
Carbon nanotubes behave as semiconductor or metal depending on the diameter and the chirality.
They opened a new field in  fundamental science 
and  have become important materials for nanotechnology devices\cite{Iijima91,DresselhausDE,SaitoDD98}. 
 Nanographene ribbons, nanoscale graphite ribbons, also show strikingly different characters depending on the shapes of their peripheral edges\cite{FujitaWNK96,WakaHari03}. The nanographene ribbon with {\it zigzag} edges (zigzag ribbon) has the peculiar nonbonding molecular orbitals  localized mainly along the zigzag edges (edge states), which cause a sharp peak in the density of states at the Fermi energy, while such states are completely absent in the ribbon with {\it armchair} edges. 
 Moreover,  zigzag ribbons will show rich natures due to the competing spin and charge orders:
 the introduction of the on-site Coulomb repulsion $U$ under the mean field approximation induces the Fermi-instability of the edge states and results in the spin-polarized  (SP) states with the opposite ferrimagnetic orders localized along the zigzag edges\cite{FujitaWNK96}; On the other hand, the nearest neighbor Coulomb repulsion $V$ induces another Fermi-instability and  results in a novel ferroelectrically charge-polarized  (CP) state with the unlike charges localized along the zigzag edges  and competes with the SP state\cite{YamaSHW}. 
 Recently, microscopic M\"{o}bius strips of $\rm{NbSe_{3}}$ were synthesized \cite{Tanda} and stimulated theoretical studies\cite{WakaHari03,HayashiE,YakuboAC}.
\begin{figure}[b]
%h=here, t=top, b=bottom, p=separate figure page
\begin{center}\leavevmode
\includegraphics[scale=0.8]{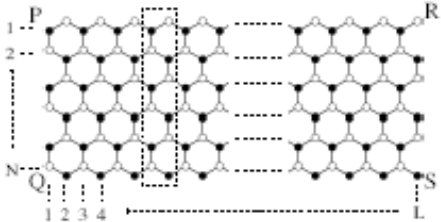}
\caption{The schematic structure of a bipartite nanographene ribbon with zigzag edges to make a non-bipartite M\"{o}bius strip or a bipartite ring; $\bullet$ ($\circ$), A (B) site in the bipartite lattice. Dashed rectangle denotes a unit cell in a ring. }
\label{figurename}
\end{center}
\end{figure}

 A M\"{o}bius strip has unique topological natures different from a ring; the former has only one side and one edge while the latter has two sides and two edges. Then, how such an exotic topology affects the electronic states in nanographene? This is the question that we study here.
 
In this paper we demonstrate that the M\"{o}bius strip has both the SP domain wall (SP-DW) state\cite{WakaHari03} and the helical state due to $U$.
 We also show that  $V$ stabilizes the novel domain wall state connecting the opposite charge orders of the CP state (CP-DW).
\section{Model and Methods}
Figure 1 illustrates the geometry of a zigzag ribbon with an inversion symmetry to make a M\"{o}bius strip or a ring.
$N$ and $L$ are the width and length of the ribbon. We consider the case both $N$ and $L$ to be even. Note that the twofold coordinated sites locate only at the zigzag edges, PR and QS, where the edge states are mainly localized. 
Since the zigzag ribbon is bipartite, 
A and B sites are assigned by black and white circles, respectively. 
The M\"{o}bius strip is obtained by twisting one end  of the ribbon, e.g., PQ  and joining the ends so that the P site is connected with the S site and the Q site is connected with the R site,
defining the M\"{o}bius boundary condition.

\begin{figure}[b]
\begin{center}
\includegraphics[scale=0.8]{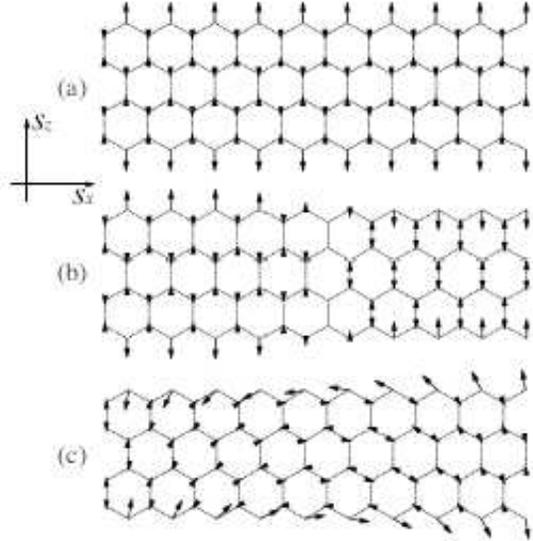}
\caption{Spin density distributions of  (a) the spin-polarized (SP) state in a ring,  (b) the SP domain wall (SP-DW) state and (c) helical states in a M\"{o}bius strip with  $N{\times}L=4{\times}20$ sites for $U=t$ and $V=0$. Arrows represent spins of $\pi$ electrons. In (a) and (b), only $z$ component of spin densities $s_{i,z}$ is finite:  In (c), $y$ component $s_{i,y}$ is zero.}
\label{figurename}\end{center}
\end{figure} 

Note that both the P and S sites belong to A sublattice 
while both the Q and R sites belong to B sublattice. Since the M\"{o}bius boundary condition forces to connect the same sublattice, the M\"{o}bius strip is not bipartite, while the periodic boundary condition keeps the system bipartite.

  We treat a half-filled  $\pi$ electron system on the zigzag ribbon  with the extended Hubbard Hamiltonian,
\begin{eqnarray}
H_{EH} = &-&t\sum_{\langle i, j \rangle, s}(c_{i,s}^{\dagger}c_{j,s} + {\rm H.c.})\nonumber\\
&+&U\sum_{i}(n_{i,\uparrow} -\frac{1}{2})(n_{i,\downarrow} -\frac{1}{2})\nonumber\\
&+&V\sum_{\langle i, j \rangle}(n_{i}- 1)(n_{j} - 1).
\end{eqnarray}
 Here, $\langle i,j\rangle$ denotes a pair of neighboring carbon sites and $-t$  is the transfer integral between them. $c^{\dagger}_{i,s}$ creates  a $\pi$ electron with spin $s$ on the $i$-th site and $c_{i,s}$ annihilates the one. $n_{i,s}$ is the corresponding number operator, and $n_{i}=\sum_{s}n_{i,s}$. 

\begin{figure}[t]
\begin{center}
\includegraphics[scale=0.8]{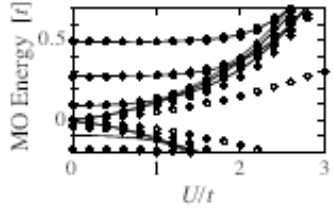}
 \caption{$U$ dependences of the MO energies of  the SP (solid line), SP-DW ($\bigcirc$) and helical ($\bullet$) states for $V=0$. }
\label{figurename}\end{center}
\end{figure}

To study the possibility of the helical state, we generalized the unrestricted Hartree-Fock approximation to set a two body operator as \cite{FujitaIN},
\begin{eqnarray}
 &\hat{c}_{i,\uparrow}^{\dagger}\hat{c}_{i,\uparrow} 
 \hat{c}_{i,\downarrow}^{\dagger}\hat{c}_{i,\downarrow}&\nonumber\\
 \equiv&\langle\hat{c}_{i,\uparrow}^{\dagger}\hat{c}_{i,\uparrow}\rangle
 \hat{c}_{i,\downarrow}^{\dagger}\hat{c}_{i,\downarrow}&
 +\hat{c}_{i,\uparrow}^{\dagger}\hat{c}_{i,\uparrow}
 \langle\hat{c}_{i,\downarrow}^{\dagger}\hat{c}_{i,\downarrow}\rangle\nonumber\\
-&\langle\hat{c}_{i,\uparrow}^{\dagger}\hat{c}_{i,\downarrow}\rangle
 \hat{c}_{i,\downarrow}^{\dagger}\hat{c}_{i,\uparrow}&
 -\hat{c}_{i,\uparrow}^{\dagger}\hat{c}_{i,\downarrow}
 \langle\hat{c}_{i,\downarrow}^{\dagger}\hat{c}_{i,\uparrow}\rangle\nonumber\\
-&\langle\hat{c}_{i,\uparrow}^{\dagger}\hat{c}_{i,\uparrow}\rangle
 \langle\hat{c}_{i,\downarrow}^{\dagger}\hat{c}_{i,\downarrow}\rangle&
 +\langle\hat{c}_{i,\uparrow}^{\dagger}\hat{c}_{i,\downarrow}\rangle
 \langle\hat{c}_{i,\downarrow}^{\dagger}\hat{c}_{i,\uparrow}\rangle.
\end{eqnarray}
The $u (=x,y,z)$ component of the spin density and the charge density at $i$-th site are given by  $s_{i,u}=\sum_{\alpha,\beta}{\langle}c_{i,\alpha}^{\dagger}({\sigma}_u)_{\alpha\beta}c_{i,\beta}{\rangle}/2$ and  $d_{i}=1-\langle n_{i}\rangle,$ respectively, where ${\sigma}_{u}$ is the Pauli matrix.
\section{Results and Discussions }
Figure 2 shows spin density distributions of (a) the SP state in a ring, and  the (b) SP-DW and (c) helical states in a M\"{o}bius strip  with $N{\times}L=4{\times}20$ carbon sites for $U=t$ and $V=0.$ 
In the SP state, the up- and down-spin ferrimagnetic orders appear along the upper  and lower zigzag edges, respectively, because the spin-density-wave like spin-orientation order is stabilized by $U$ in the bipartite system with the edge states at twofold sites \cite{FujitaWNK96}. The M\"{o}bius boundary condition forces to connect the opposite ferrimagnetic orders along the zigzag edge and causes the magnetic frustration. It allows the SP-DW state with the opposite ferrimagnetic orders connected by the interface where the magnitudes of spins are suppressed to reduce the frustration\cite{WakaHari03}. On the other hand, the helical state keeps the magnitudes of spins along the zigzag edge with the ferrimagnetic feature in a unit cell. In this state, the directions of spins at twofold sites rotate uniformly along the zigzag edge to circumvent the magnetic frustration due to the M\"{o}bius boundary condition. The spin rotates by $2\pi$ in the $s_{x}s_{z}$ plane during the travel along the zigzag edge to return to the starting point, showing the helical ferrimagnetic order. Therefore, the SP-DW and helical states satisfy the constraint of the M\"{o}bius boundary condition connecting the opposite zigzag edges of the ribbon, i.e., the opposite ferrimagnetic domains in the SP state. A transformation  of all spins in the helical state under an Euler angle rotation in the spin space causes another degenerate helical state, since the spin space is isotropic in this model. Total  magnetization are zero in the SP-DW and helical states.

\begin{figure}[t]
\begin{center}
\includegraphics[scale=0.8]{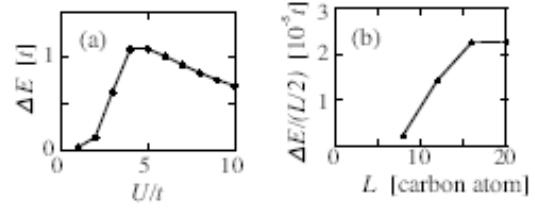}
 \caption{(a) $U$ dependence of the energy difference ${\Delta}E=E_{\rm SP-DW}-E_{\rm helical}$ for the system with $4{\times}40$ sites and $V=0,$ where $E_{\rm SP-DW}$ and $E_{\rm helical}$ are the energies of the SP-DW and helical states, respectively. (b) Dependence of ${\Delta}E$  per unit cell on the ribbon length $L$ for $U=t, V=0$ and $N=4$.}
\label{figurename}\end{center}
\end{figure} 

In the case of much larger $U$ (${\sim}10t$), 
the magnitudes of spin densities become almost the same at all sites in the SP and  helical states as well as in each ferrimagnetic domain of the SP-DW state: the ferrimagnetic feature turns into the antiferromagnetic one with increasing $U$.

Figure 3 shows the $U$ dependences of the molecular orbital (MO) energies around the Fermi level in the SP,  SP-DW and helical states for the system with $N{\times}L=4{\times}40$ sites. In the SP state, $U$ opens the energy gap between the degenerate edge states at the Fermi energy, while in the SP-DW state the frontier orbitals remain inside the gap \cite{WakaHari03} similar to the soliton state in polyacetylene \cite{Heeger}. On the other hand, in  the helical state the frontier orbitals behave more similarly to that in the SP state than in the SP-DW state. This is because the variation of the spin direction in the helical state is extended over the system.  In the shorter ribbon with $N{\times}L=4{\times}20,$ the difference between their behaviors in the SP-DW and helical states is not so appreciable because a half pitch $L$ in the helical state is closer to the size of the interface in the SP-DW state.

 For $U>0$, the helical state is  lower in energy than the SP-DW state, as shown in Fig. 4 (a). This is consistent with the MO's behavior  shown in Fig. 3: the wider gap and the enhanced decrease in the occupied MO energies in the helical states than in the SP-DW states. 
 Figure 4 (b) shows that the energy difference per unit cell is smaller in the shorter ribbon, as $L$ approaches the size of the interface in the SP-DW state.

We recently reported that $V$ stabilizes the novel CP state with a finite electric dipole moment under periodic boundary condition\cite{YamaSHW}. Figure 5 (a) shows the charge density distribution of this state for $U=0.2t$ and $V=0.4t.$ The positive and negative charges are localized along the lower and upper zigzag edges, respectively, due to $V$ and they form a finite electric dipole moment pointing from the upper edge to the lower edge. Net charge along each zigzag edge arises from much larger charge density at twofold sites than at threefold sites due to the edge states. The magnitude of the electric dipole moment per unit cell is 6.2 Debye in the parameters. This exotic electric dipole moment is induced by $V$  and breaks the inversion symmetry of the system. The CP state competes with the SP state.

\begin{figure}[t]
\begin{center}
\includegraphics[scale=0.8]{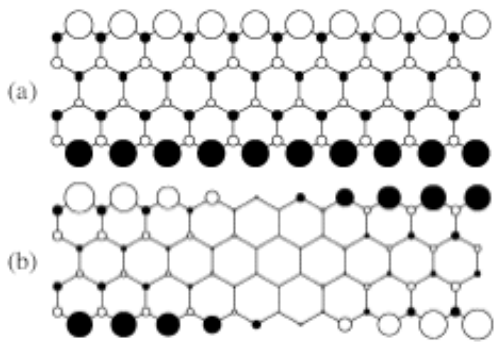}
 \caption{Charge density, $d_{i}$, distribution of (a) the charge-polarized (CP) state in a zigzag ring and (b) the domain wall state connecting the opposite charge orders of the CP state in a M\"{o}bius strip for $U=0.2t$ and $V=0.4t$; $\bullet (\circ)$, positive (negative) charge densities. The radius of each circle denotes the magnitude of the charge density; the maximal ones are 0.23 in (a)  and 0.24 in (b). The system sizes are $N{\times}L=4{\times}20$ sites.}
\label{figurename}\end{center}
\end{figure} 

In a M\"{o}bius strip, corresponding domain wall state, CP-DW state, is stabilized as shown in Fig. 5 (b). It connects the opposite charge orders of the CP state. The M\"{o}bius boundary condition allows only the CP-DW state  in this case since charge is a scalar, while it allows the SP-DW and helical states in the spin ordered case since spin is a vector. The phase diagram for their competition in the $U$-$V$ space will be reported elsewhere.
\section{Conclusion}
In summary, we demonstrated that a M\"{o}bius strip of nanographene with a zigzag edge has the spin-polarized domain wall state and the helical ferrimagnetic state caused by the on-site Coulomb repulsion within the extended-Hubbard model under the mean field approximation and the former is lower in energy than the latter. It also has the charge-polarized domain wall state stabilized by the nearest neighbor Coulomb repulsion which connects the opposite charge orders.

\section*{Acknowledgements}
This work has been supported partly by Special Coordination Funds for Promoting Science and Technology and by NEDO under the Nanotechnology Materials Program. One of us (A. Y.) thanks all the members in his laboratory for helpful discussions and supports.

\end{document}